\documentclass[useAMS,usenatbib]{mnras}

\usepackage{tabto}
\usepackage{float}
\usepackage{color}
\usepackage{amssymb} 
\usepackage{amsfonts,amsmath,amscd} 
\usepackage[subnum]{cases}
\usepackage{mathtools}
\usepackage{verbatim}
\usepackage{t1enc}
\usepackage[T1]{fontenc}
\usepackage{ae,aecompl}
\usepackage[dvips]{graphics}
\usepackage{adjustbox}
\usepackage{eufrak} 
\usepackage{euscript} 
\usepackage{graphicx} 
\usepackage[normalem]{ulem} 
\usepackage{makeidx} 
\usepackage[usenames,dvipsnames]{pstricks}
\usepackage{epsfig}
\usepackage{pst-grad} 
\usepackage{pst-plot} 
\usepackage{pstcol,times,graphics,pstricks,pst-node,pst-coil} 
\usepackage{multicol} 
\usepackage{multirow} 
\usepackage{indentfirst} 
\usepackage{natbib} 
\usepackage{setspace} 
\usepackage{threeparttable}

\title[Dynamical equivalence, MW pop. and clusters]
{Dynamical equivalence, the origin of the Galactic field stellar and binary population, and the initial radius--mass relation of embedded clusters}
\author[D. Belloni et al.]
{Diogo Belloni$^{1,2}$\thanks{E-mail: belloni@camk.edu.pl},
Pavel Kroupa$^{3,4}$, Helio J. Rocha-Pinto$^5$ and Mirek Giersz$^{1}$\\
$^{1}$ Nicolaus Copernicus Astronomical Centre, Polish Academy of Sciences, ul. Bartycka 18, PL-00-716 Warsaw, Poland \\
$^{2}$ CAPES Foundation, Ministry of Education of Brazil, DF 70040-020, Brasilia, Brazil \\
$^{3}$ Helmholtz-Institut f\"ur Strahlen- und Kernphysik, Nussallee 1416, D-53115 Bonn, Germany \\
$^{4}$ Charles University in Prague, Faculty of Mathematics and Physics, Astronomical Institute, V Holesovickach 2, CZ-180 00 Praha 8, Czech Republic \\
$^{5}$ Universidade Federal do Rio de Janeiro, Observat{\'o}rio do Valongo, Ladeira do Pedro Ant\^onio 43, 20080-090 Rio de Janeiro, Brazil}
\begin{document}

\date{ Accepted 2017 November 20. Received 2017 November 18; in original form 2017 October 7}

\pagerange{\pageref{firstpage}--\pageref{lastpage}} \pubyear{2016}

\maketitle

\label{firstpage}

\begin{abstract}
In order to allow a better understanding of the origin of Galactic field populations, dynamical 
equivalence of stellar-dynamical systems has been postulated by Kroupa and Belloni et al. 
to allow mapping of solutions of the initial conditions of embedded clusters such that they yield, 
after a period of dynamical processing, the Galactic field population. Dynamically equivalent systems are defined 
to initially and finally have the same distribution functions of periods, mass ratios and eccentricities 
of binary stars. Here we search for dynamically equivalent clusters using the {\sc mocca} code. 
The simulations confirm that dynamically equivalent solutions indeed exist. The result is that the 
solution space is next to identical to the radius--mass relation of Marks \& Kroupa, 
$\left( r_h/{\rm pc} \right)= 0.1^{+0.07}_{-0.04}\, \left( M_{\rm ecl}/{\rm M}_\odot \right)^{0.13\pm0.04}$.
This relation is in good agreement with the observed density of molecular cloud clumps. According to the solutions, 
the time-scale to reach dynamical equivalence is about 0.5~Myr which is, interestingly, consistent 
with the lifetime of ultra-compact HII regions and the time-scale needed for gas expulsion to be 
active in observed very young clusters as based on their dynamical modelling. 
\end{abstract}

\begin{keywords}
methods: numerical -- binaries: general -- open clusters and associations: general -- star clusters: general
\end{keywords}

\section{INTRODUCTION}

The Milky Way field population of stars is understood to come from the
dissolution of embedded star clusters after the expulsion of the residual gas in the star formation
process. Such a clustered star formation, followed by the dissolution 
of these embedded clusters, is expected to be the dominant process 
that populates the Galactic field with stars 
\citep[e.g.][]{Lada_2003,Lada_2010,Kroupa_2011,Marks_2011,Megeath_2016}.

In order to bring into agreement observations of pre-main-sequence and
Galactic field population stars, \citet{Kroupa_1995a,Kroupa_1995b,Kroupa_1995c}
derived the birth (or initial) distribution functions of late-type binary systems and
developed a simple model for the redistribution
of energy and angular momentum in proto-binary systems such that
it directly leads to observed short-period main-sequence binary correlations
(mass ratio, eccentricity and period), which is known as pre-main-sequence eigenevolution.
After pre-main-sequence eigenevolution, dynamical evolution of these
embedded clusters is expected to change the properties of the binaries
(mainly long-period ones) such that, after the gas removal and expansion of the
clusters, the Milky Way is populated with single and binary stars with
properties similar to those observed.

Such a model has been checked against numerical simulations
and observations and has successfully explained observational features 
of young clusters, associations, the Galactic field and even
binaries in old globular clusters 
\citep[e.g][and references therein]{Kroupa_2011,Marks_2012,Leigh_2015,Belloni_2017b}.

Particularly interesting is that,
assuming the cluster origin of the field population, leads to theoretical 
semi-major axis distributions that are consistent with the observed ones
in six young clusters and star-forming regions, as well as in two
older open clusters \citep{Marks_2012}. For all investigated objects, the theoretical
semi-major axis distributions turn out to be parent functions of the
observational data. In addition, these authors found a weak half-mass radius--stellar mass
correlation for cluster-forming cloud clumps of the form
$\left( r_h/{\rm pc} \right)\,=\,0.1^{+0.07}_{-0.04} \; \left( M_{\rm ecl}/{\rm M_\odot}\right)^{0.13 \pm 0.04}$.
These results suggest that the initial binary properties do not vary 
significantly between different environments within the Galaxy.
In addition, the observed surface density distribution of very young stars is 
consistent with all of them stemming from compact embedded clusters \citep{Gieles_2012}.


Here we test this model by looking for the initial cluster conditions such that 
the Milky Way field population can be reproduced, after dynamical processing during 
the embedded phase, and look for the predicted radius--mass relation in order to
compare with that derived from observations. 

A crucial concept used here is {\it dynamical equivalence}, 
which can be defined as follows
\citep[][]{Kroupa_1995a,Kroupa_1995b,Kroupa_1995c,Belloni_2017b}: 
if two clusters with different masses and different initial radii dynamically evolve an
identical initial binary population to similar distribution functions of 
binaries, then these two clusters are `dynamically equivalent'. A key part of this
concept is that cluster evolution time-scales are not necessarily the same.

Our aim in this work is to search for the set of dynamically equivalent models, i.e. 
to find the combinations of mass, radius and time, which are able to reproduce observed 
properties of stellar populations in the Galaxy and compare it with observations of embedded clusters.


\section{CLUSTER SIMULATIONS}
\label{simulations}

For our simulations, we have used the MOnte-Carlo Cluster simulAtor
({\sc mocca}) code developed by \citet[][and references therein]{Giersz_2013},
which includes the {\sc fewbody} code \citep{Fregeau_2004} to perform 
numerical scattering experiments of small-number gravitational interactions and
the {\sc sse/bse} code \citep{Hurley_2000,Hurley_2002} to 
deal with effects of both single and binary stellar evolution.
{\sc mocca} has been extensively tested against $N$-body codes 
\citep[e.g.][]{Giersz_2013,Wang_2016,Madrid_2017} and 
reproduces $N$-body results with good precision, not 
only for the rate of cluster evolution and the cluster mass distribution 
but also for the detailed distributions of mass and binding energy of 
binaries.

The initial cluster models are assumed to be spherically symmetric and virialized,
and they have neither rotation nor mass segregation.
For all simulations, we have adopted the \citet{Kroupa_2001} canonical
stellar initial mass function (IMF) with stellar masses ranging from 0.08 M$_\odot$
to $M_{\rm max}$, where $M_{\rm max}$ is the maximum stellar mass allowed
to form in a cluster with mass $M_{\rm ecl}$ \citep[][]{Weidner_2013}.
All models have 95 per cent of binaries, a King density profile ($W_0 = 6$) 
and adopt the \citet{Kroupa_1995b} initial binary population, with the 
improvements described in \citet{Belloni_2017b}.

In what follows, we briefly describe the initial binary population adopted
in this work, whose detailed description can be found in \citet{Belloni_2017b}.
For consistency, pre-main-sequence eigenevolution is applied only to 
binaries whose primaries are less massive than 5 M$_\odot$, which is reasonable
because the time-scale of pre-main-sequence evolution of massive stars is short 
enough that it is safe to neglect it \citep{Railton_2014}. In this case,
the initial population of massive binaries is assumed to be identical to their
birth population, which is assumed to be that inferred by \citet{Sana_2012}, with
the normalizations described in \citet{Belloni_2017b}. In particular, the
mass ratio distribution is flat and the period and eccentricity distributions are 
such that they favour short-period binaries and binaries with small eccentricity.

The birth population associated with low-mass binaries has the following properties:
(i) all star masses are randomly chosen from the Kroupa canonical IMF; 
(ii) the binaries are created by randomly pairing the stars, from the list made in step (i);
(iii) the eccentricity distribution is thermal; and
(iv) the period distribution follows Eq. 8 in \citet{Kroupa_1995b}.
After the birth population is generated, we apply pre-main-sequence eigenevolution, 
as described in \citet{Belloni_2017b},
in order to generate the initial distributions.

Finally, we stress that the best way of pairing birth low-mass and high-mass binaries is by preserving
the IMF. This is achieved by applying a similar procedure as described in section 6.3 of 
\citet[][see also \citet{Oh_2015,Oh_2016}]{Belloni_2017b}, 
i.e. we first generate an array of all stars and after that we pair the stars in a way consistent with
either a uniform distribution (high-mass binaries) or random pairing (low-mass binaries).

In order to look for the class of dynamically equivalent solutions
that is able to reproduce the Galactic field late-type binary properties,
we set a grid of models varying the number of objects (single stars plus binaries)
and half-mass radius ($r_h$), being $N_{\rm obj} \in [10^4;10^5]$, 
in steps of $10^4$ and $r_h \in [0.1;1.5]$ pc, in steps of
$0.05$ pc. We stress that in our simulations, setting the number of objects
is equivalent to setting the cluster mass, since the IMF is the same in all cases. 
In addition, the lower limit for $N_{\rm obj}$ is because {\sc mocca} simulations are reliable
only when a model has at least $10^4$ objects, while the upper limit
is justified by the maximum cluster mass from which the Galactic field
originates, which is $\approx 10^5$ M$_\odot$ \citep{Marks_2011b}.

Each combination of $(M_{\rm ecl},r_h)$ was repeated 20 times with a different initial random 
number seed to account for model stochasticity. Finally, all clusters were evolved for three different time-scales
(duration of the dynamical evolution), namely $t \approx $ 0.5, 1.0 and 3.0~Myr, 
which is assumed here to be the time of residual gas removal (e.g. due to the evolution
of the most massive stars in the clusters)
\citep[e.g.][]{Kroupa_2011,Banerjee_2013,Banerjee_2014,Brinkmann_2017}.

For each mass in the grid, we search for the best-fitting model (i.e. the best $r_h$)
such that, after the dynamical evolution, binary distributions are similar
to those observed in the Galactic field result. To find the best model for each
cluster mass, we used properties of late-type binaries to compare with observational 
data of G-dwarf binaries in the Galaxy \citep[][]{DM_1991}, using the minimum 
$\chi^2$ method of model fitting. 

We took as the best-fitting combination of $(M_{\rm ecl},r_h)$ that corresponding to the minimum $\chi^2$. The estimated errors for this best-fitting combination were found from the standard deviation of the $M_{\rm ecl}$ and $r_h$ values of those models yielding the 20 per cent lowest $\chi^2$ statistics\footnote{We have found that the $\chi^2$ surface in the $(M_{\rm ecl},r_h)$ plane does not present a clear and well-separated global minimum for this problem. This is why we estimate the error in $M_{\rm ecl}$ and $r_h$ values using an arbitrary cut at the 20\% lowest $\chi^2$. Indeed, we have analysed how this error estimate is affected by the maximum $\chi^2$ cut. We have found that this error estimate is nearly constant when we cut the sample at the 20\%\,--\,40\% lowest $\chi^2$. We decided to use the minimum value of this range, 20\%, to avoid averaging over $(M_{\rm ecl},r_h)$ much different from the best-fitting values. }.

We stress that our aim is to find
the combinations of $M_{\rm ecl}$ and $r_h$ that lead to binary distributions
similar to those observed. In other words, our procedure provides, in the end, a set of 
models in the plane $(M_{\rm ecl},r_h)$ that not only have binary properties, 
after dynamical evolution, similar to those observed in the field, but are also
dynamically equivalent.

\begin{figure*}
   \begin{center}
    \includegraphics[width=0.4\linewidth]{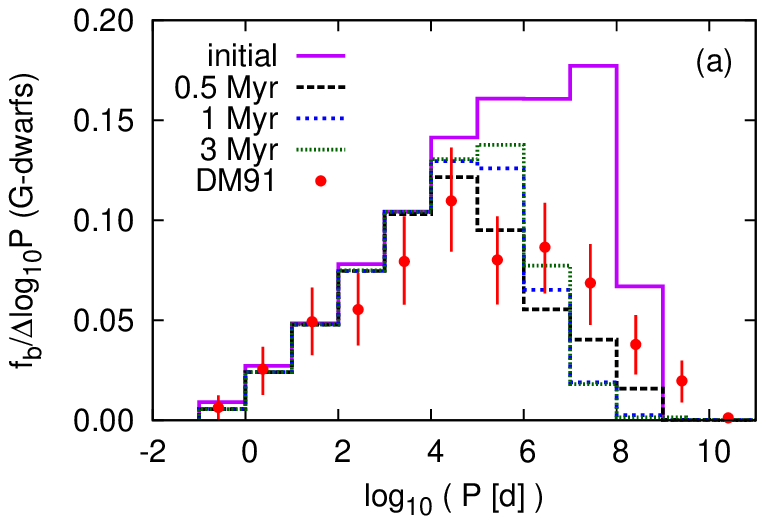} \hspace{1.0cm}
    \includegraphics[width=0.4\linewidth]{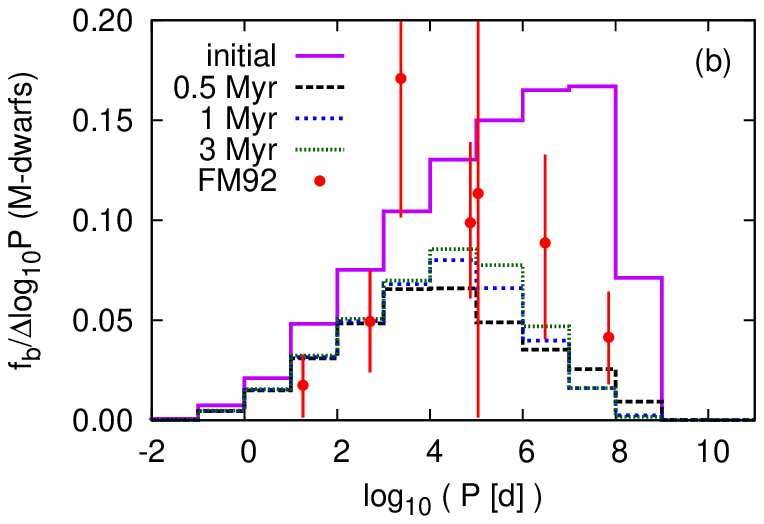} 
    \includegraphics[width=0.4\linewidth]{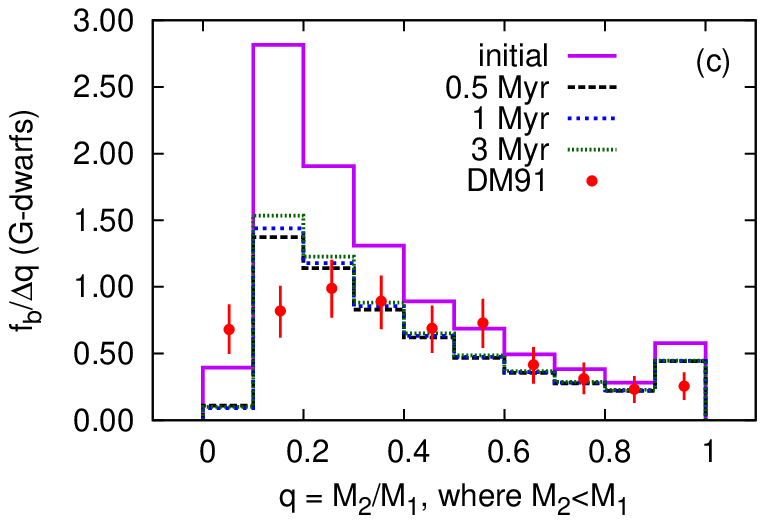} \hspace{1.0cm} 
    \includegraphics[width=0.4\linewidth]{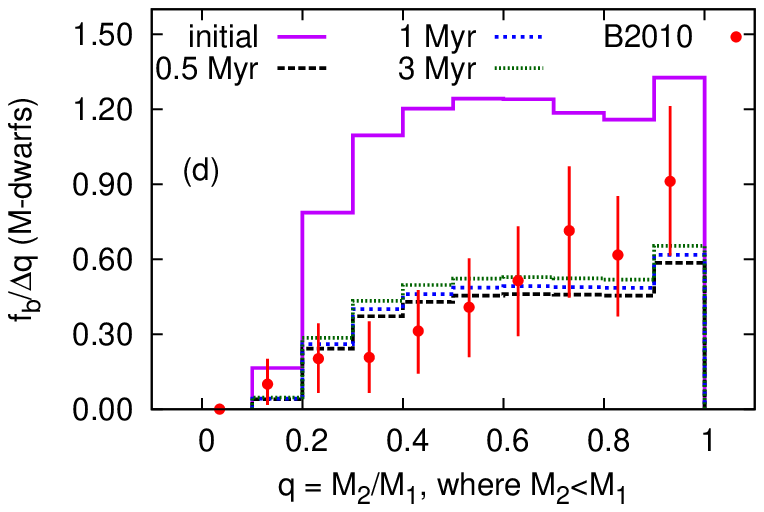} 
    \includegraphics[width=0.4\linewidth]{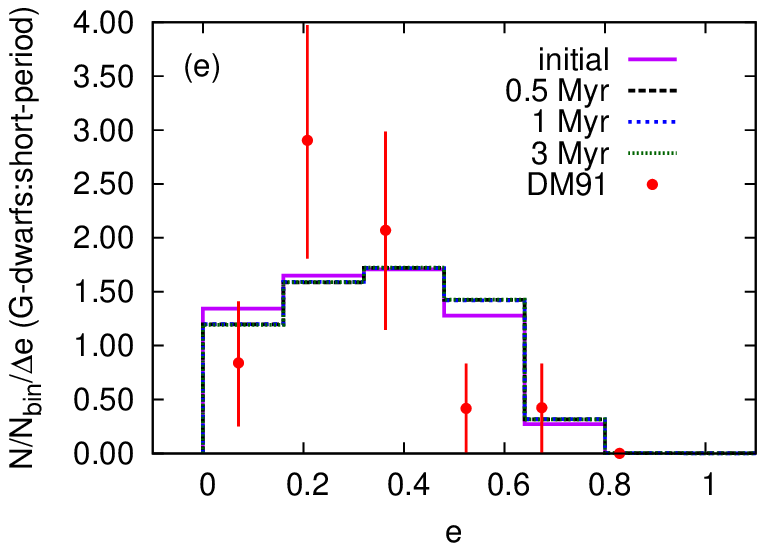} \hspace{1.0cm} 
    \includegraphics[width=0.4\linewidth]{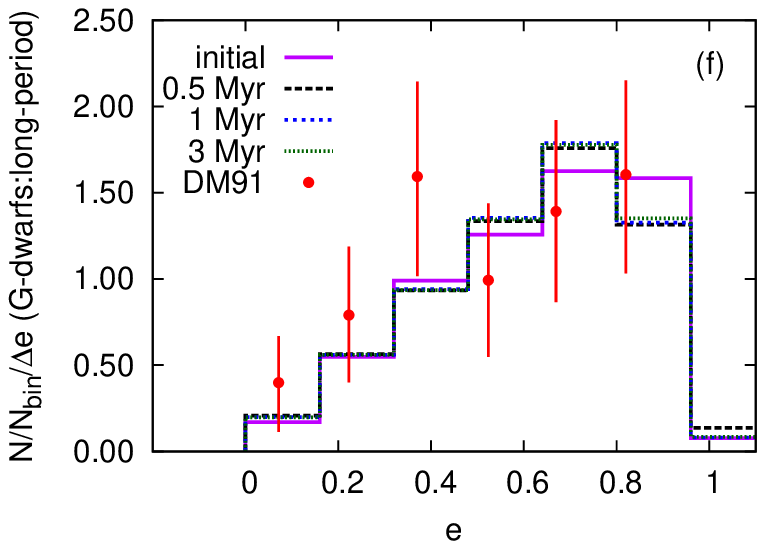} 
    \includegraphics[width=0.4\linewidth]{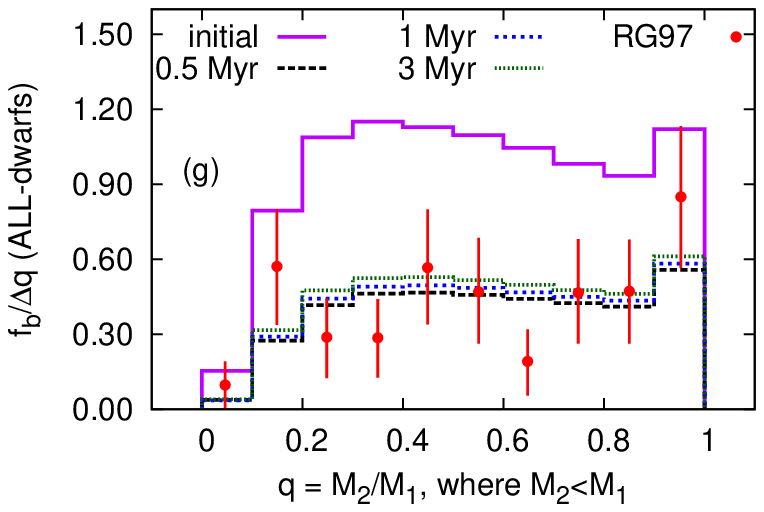} \hspace{1.0cm}
    \includegraphics[width=0.4\linewidth]{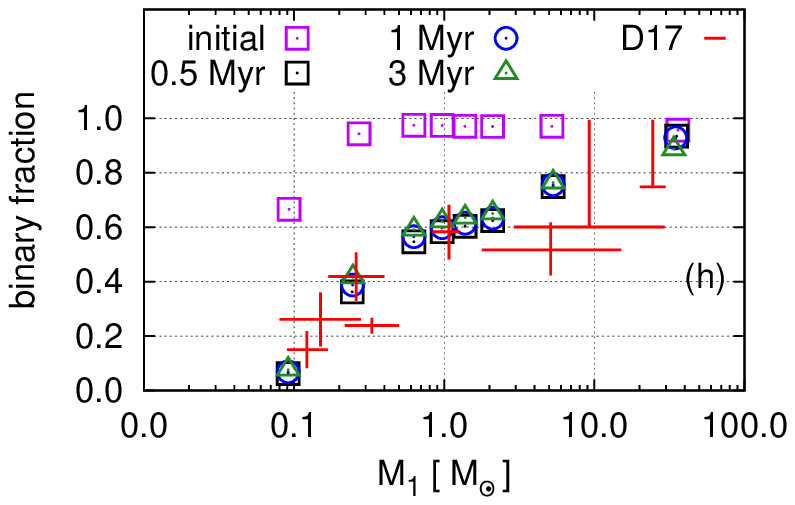} 
   \end{center}
  \caption{Predicted and observed Galactic binary properties. 
In panels (a) and (b) are plotted the period distributions of G- and M-dwarfs, respectively. 
In panels (c) and (d) we show the mass ratio distributions of G- and M-dwarfs, respectively.
In panels (e) and (f), we exhibit the eccentricity distributions for short-period and long-period G-dwarfs, respectively.
Mass ratio distributions for all dwarfs such that $M_1$ is smaller than 2 M$_\odot$ are shown in panel (g).
Finally, in panel (h), we show the predicted and observed primary mass-dependent binary fraction
(see also \citet{Thies_2015,Marks_2017}). 
Observational distributions in panels (a)--(g) are plotted with red filled circles and
were extracted from \citet{DM_1991} (DM91), \citet{M_1992} (M1992), \citet{FM_1992} (FM92), 
\citet{RG_1997} (RG97), and \citet{B_2010} (B2010). 
We normalized the distributions such that the area under each distribution equals the total binary fraction in the
Galactic field, with the exception of panels (e) and (f), where we normalized with respect to the total number of binaries,
which is needed for a consistent comparison with the observational data.
The observational binary fractions
were extracted from \citet{Dorval_2017} (D17), with the exception of Solar-like stars whose
binary fraction was extracted from DM91.
The predicted properties correspond to all dynamically equivalent
solutions taking into account the 20 per cent lowest $\chi^2$ statistics in the fitting procedure for the
three time-scales for dynamical evolution adopted here. 
Notice that predicted and observed distributions are remarkably compatible with each other.
For more details, see Section \ref{results}.}
  \label{Fig01}
\end{figure*}


\section{RESULTS AND DISCUSSION}
\label{results}

In Fig. \ref{Fig01} we compare our best-fitting models with observational
data of Milky Way binaries. In the figure we included all combinations of $(M_{\rm ecl},r_h)$, taking only
those models yielding the 20 per cent lowest $\chi^2$ statistics for each $M_{\rm ecl}$. 
We also included in the figure the initial distributions (i.e. at $t=0$, after pre-main-sequence eigenevolution) so that readers can
assimilate how significant are the changes in the binary properties due to dynamical evolution.
Similarly to what we did in \citet{Belloni_2017b}, we have used distributions extracted from the following 
studies\footnote{Our definitions for the late-type binaries are as follows: all late-type, G and M-dwarfs have primary
masses in the ranges $[0.08,2.0]$, $[0.8,1.2]$, and $[0.08,0.6]$, respectively, all in units of M$_\odot$.}:
(i) G-dwarf period distribution:  \citet[][hereafter DM91]{DM_1991};
(ii) M-dwarf period distribution:  \citet[][hereafter FM92]{FM_1992};
(iii) G-dwarf eccentricity distribution: DM91;
(iv) G-dwarf mass ratio distribution:  DM91 and \citet[][hereafter M1992]{M_1992};
(v)  M-dwarf mass ratio distribution:  \citet[][hereafter B2010]{B_2010};
(vi) late-type-dwarf mass ratio distribution:  \citet[][hereafter RG97]{RG_1997}.
For the binary fraction of late- and early-type stars, 
we have used the compilation by \citet[][hereafter D17]{Dorval_2017},
with the exception of G-dwarf binaries whose binary fraction was extracted from DM91.
Notice that our best-fitting models, for different time-scales of dynamical evolution,
agree well with observations.

\begin{figure}
  \begin{center}
    \includegraphics[width=0.95\linewidth]{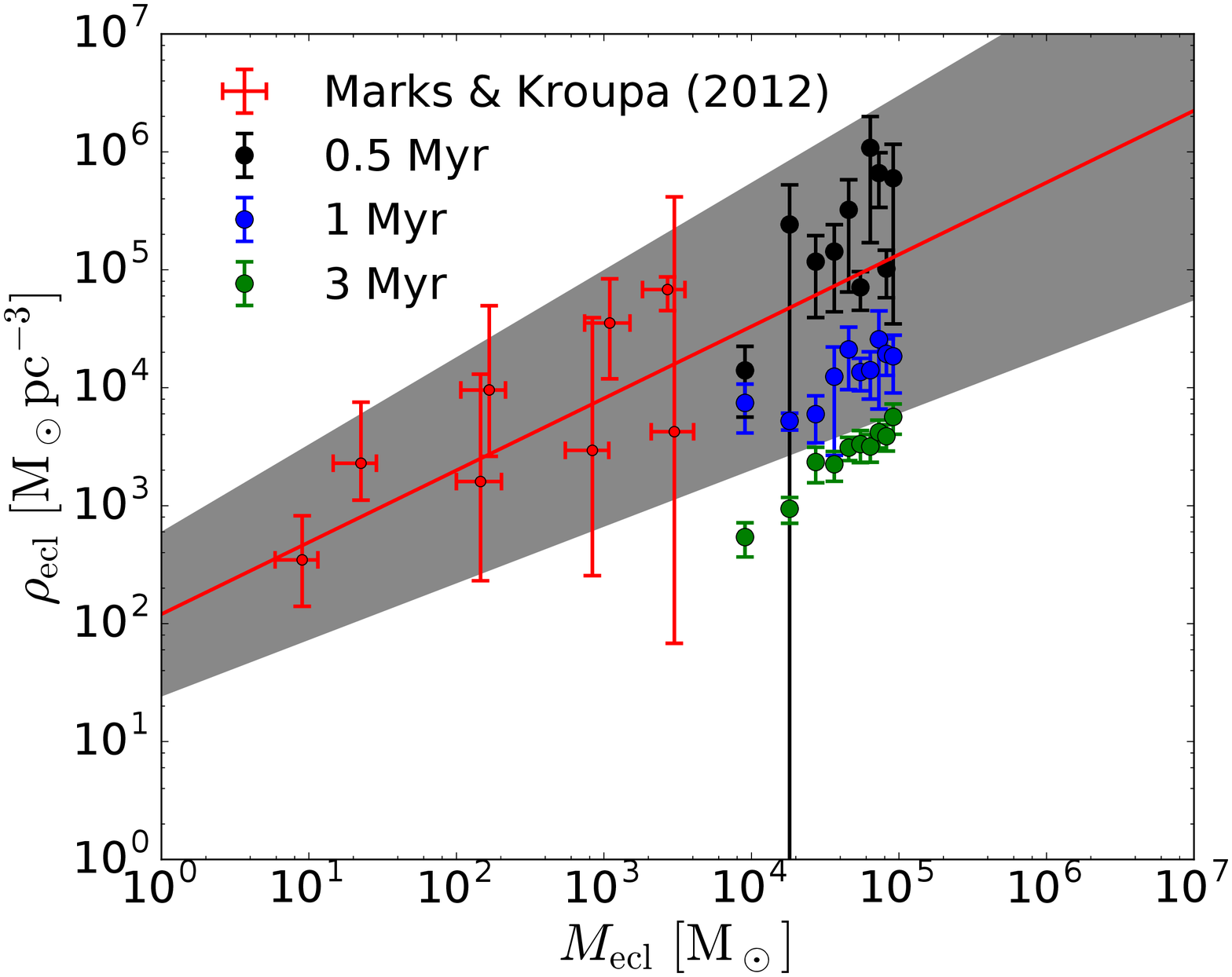} 
    \includegraphics[width=0.95\linewidth]{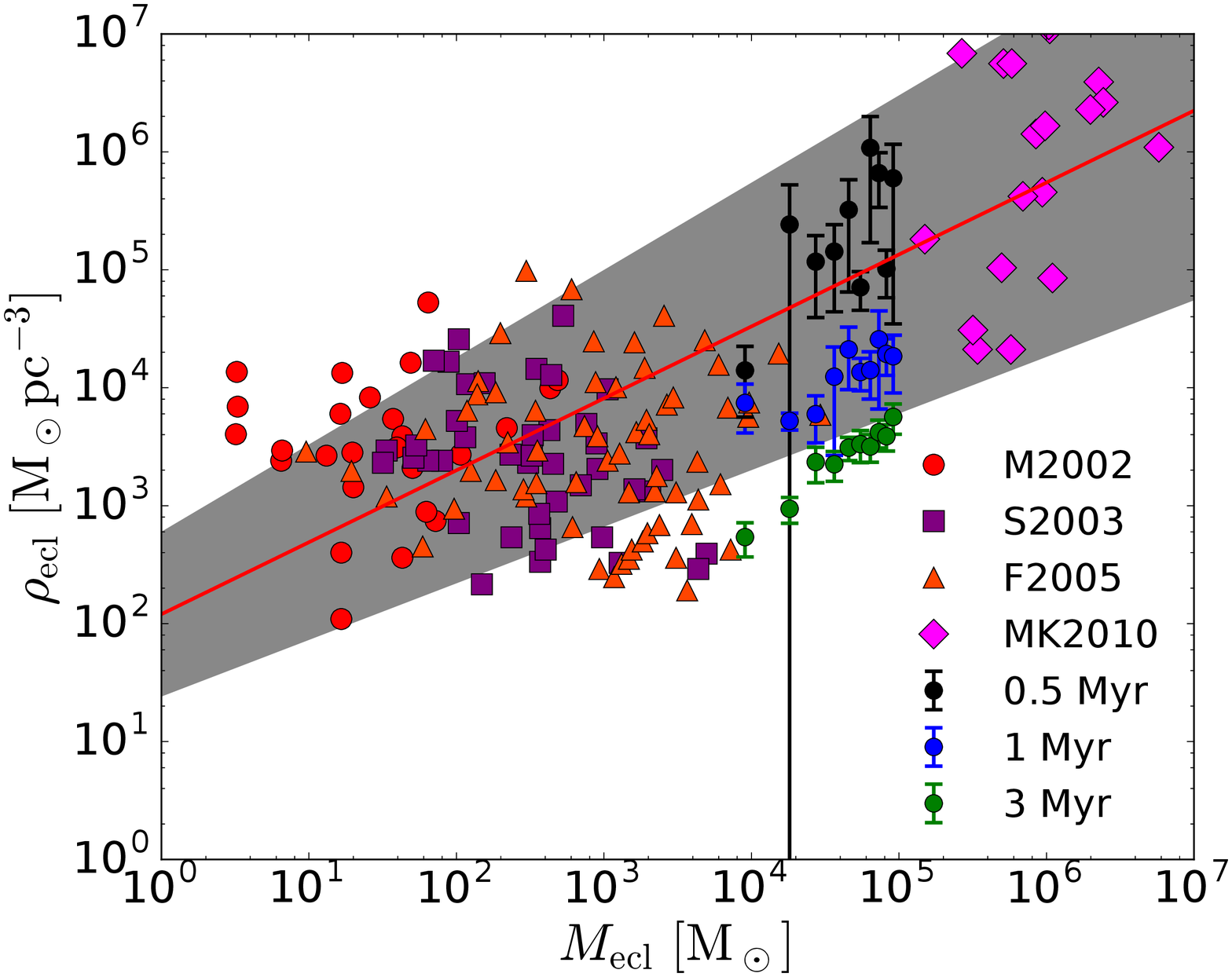} 
    \includegraphics[width=0.95\linewidth]{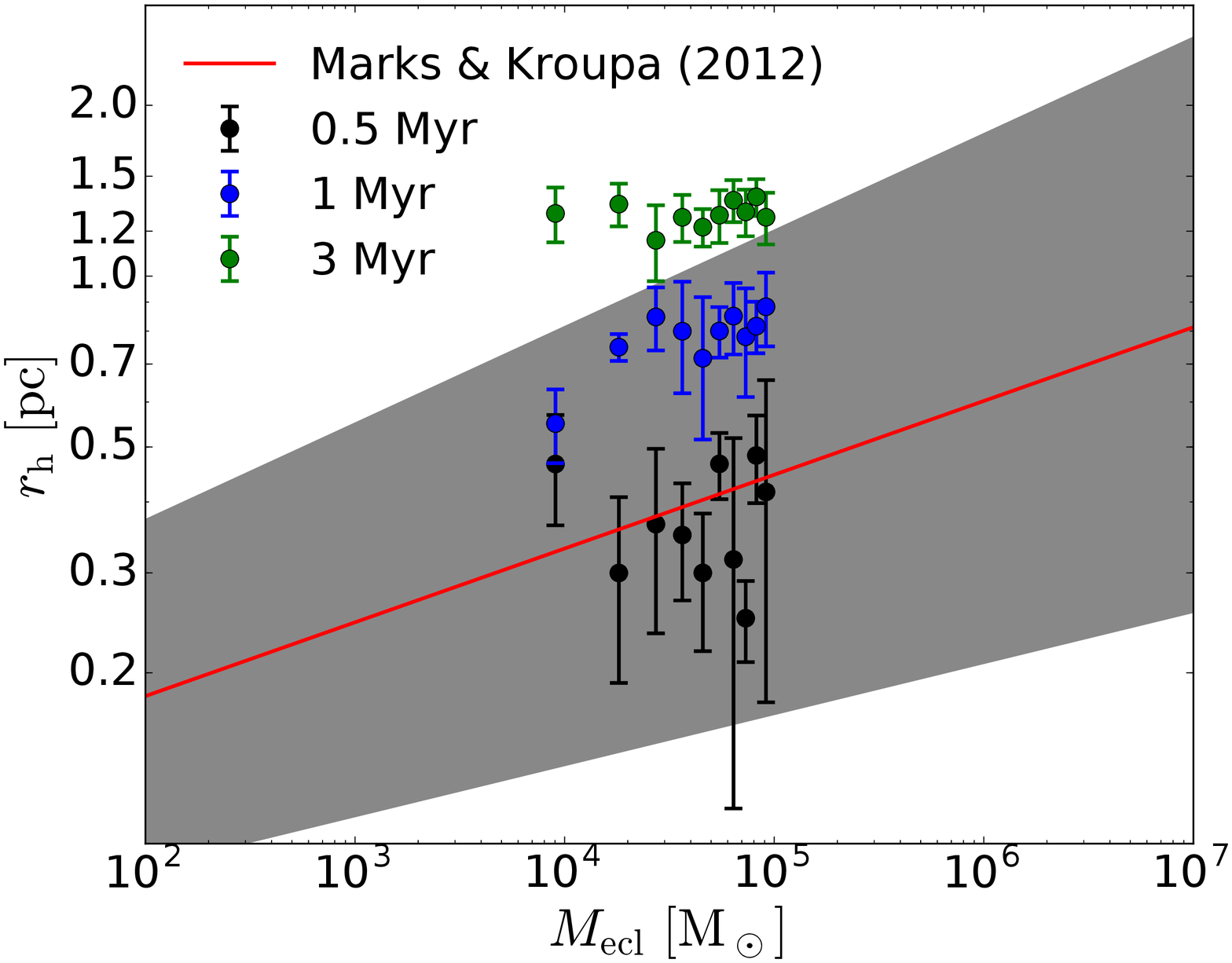} 
  \end{center}
  \caption{Initial cluster properties. Shown are the cluster mass vs. average density within the half-mass radius in the top and middle panels,
and the cluster mass vs. half-mass radius in the bottom panel. 
The observational data were taken from 
\citet{Marks_2012}, \citet{Mueller_2002} (M2002), \citet{Shirley_2003} (S2003), \citet{Fontani_2005} (F2005), and \citet{Marks_2010} (MK2010), 
and correspond to molecular cloud clumps, star-forming regions and globular clusters. 
The red line corresponds to the
correlation obtained by \citet{Marks_2012} and the grey area is the band of $1\sigma$ around it. 
The errors associated with the best-fitting models in this work were found from the standard deviation of 
the $M_{\rm ecl}$ and $r_h$ values of those models yielding the 20 per cent lowest $\chi^2$ statistics. 
For more details, see Section \ref{results}.}
  \label{Fig02}
\end{figure}

The first thing we notice in Fig. \ref{Fig01} is that we have such a good agreement, even
considering different time-scales for dynamical evolution, when comparing predicted and
observed Galactic field binary properties.
In particular, the agreement is achieved for different measures of the transformation of the
initial to final distributions, namely for the period (or semi-major axis), the mass ratio
and the eccentricity distributions.
This is directly connected with the concept of dynamical equivalence.
Given that dynamically equivalent models evolve their binaries similarly,
we can confirm that the main parameters behind this concept are the initial cluster mass,
initial cluster density (or initial half-mass radius) and duration of the dynamical processing.
As an example, consider a cluster whose initial mass is $10^4$ M$_\odot$. Let us then
consider three initial half-mass radii, namely 0.3, 0.5 and 0.7 pc. One behaviour we might expect
is that the cluster with $r_h=0.7$ pc will require a longer dynamical evolution than the models
with $r_h=0.5$ or $0.3$ pc. This is because the cluster with $r_h=0.7$ pc is less dense than the
others, given that they all have the same mass, and that the probability of interactions
depends on the cluster density, as also quantified by \citet{Marks_2012}. 
In this way, this cluster will require more time to significantly 
change its binary population by dynamical interactions.
In a similar way, the cluster with $r_h=0.5$ pc needs a shorter dynamical evolution than the model with
$r_h=0.7$ pc, but a longer evolution than the model with $r_h=0.3$ pc. 

This simple exercise allows us 
to properly see the advantage of the concept of dynamical equivalence: for a given mass, 
the denser the cluster, the quicker is the dynamical evolution. An alternative way of stating this
is: for a given duration of the dynamical evolution, the greater the mass, the larger
the cluster radius needs to be for dynamical equivalence. 

These two ways of understanding dynamical equivalence are clearly illustrated in Fig. \ref{Fig02},
where we show the initial radius--mass and initial density--mass relations for the three
different dynamical evolution time-scales adopted here. The observational data are for
(i) young clusters and star-formation regions: \citet{Marks_2012};
(ii) molecular cloud clumps: \citet[][hereafter M2002]{Mueller_2002}, \citet[][hereafter S2003]{Shirley_2003}, and \citet[][hereafter F2005]{Fontani_2005};
(iii) globular clusters: \citet[][hereafter MK2010]{Marks_2010}. The red line is the correlation obtained
by \citet{Marks_2012} and the grey area corresponds to the $1\sigma$ error in the fitting process 
and was also determined by \citet{Marks_2012}.

Notice that in the plane $\rho_{\rm ecl}$ vs. $M_{\rm ecl}$, a correlation is clearly evident
when considering all best-fitting models, when fixing the duration of the dynamical evolution
(which is separated in the figure with different colours). 
Indeed, we carried out Pearson's rank correlation tests, and we found a strong correlation
with at least 96.9 per cent confidence, in all cases.
This correlation is associated
with dynamical equivalence which allows us to always find initial cluster conditions together
with different durations of dynamical evolution such that, after such an evolution, the
properties of the Galactic field binaries result. In the particular case of the
three time-scales adopted here, we notice that the shorter the duration of the dynamical
evolution is allowed to be, the denser the cluster needs to be. 

Now, comparing in the same plane with properties of embedded clusters constrained from
observations, we see that a time-scale of only $\lesssim$ 1~Myr is needed to process dynamically 
the initial binary population.
In fact, notice that dynamically equivalent models evolved
for $\approx$ 0.5 and 1~Myr lie within the $1\sigma$ error associated with observed clusters.
This indicates that this is the time-scale needed to process dynamically the initial binaries,
to reproduce the binary properties of the Galactic field population. In other words, our results suggest that
the time-scale for gas removal is $\lesssim$ 1~Myr.
This is in very good agreement with
the properties of massive star-burst clusters 
\citep[][and references therein]{Kroupa_2001b,Banerjee_2015,Banerjee_2017,Brinkmann_2017},
e.g. NGC 3603, which suggest this time-scale during which gas is still significant 
in clusters.
It is also about the time-scale on which the formation of the embedded
cluster takes place and ultra-compact HII regions are estimated to last and
to break out to become HII regions 
\citep{Wood_1989,Churchwell_2002,Churchwell_2010}.

Notice that the above conclusions are evident also from the plane
$r_h$ vs. $M_{\rm ecl}$.
Dynamically equivalent models evolved for shorter time-scales have
smaller radii, for a given mass. In particular, models evolved for
0.5~Myr have radii smaller than models evolved for 1~Myr and 3~Myr,
and models evolved for 1~Myr have radii smaller than those evolved
for 3~Myr.
In addition, we notice that only dynamically equivalent models evolved for $\lesssim$ 1~Myr are
within the uncertainty of the radius--mass relation inferred from
observations, being $\approx$ 0.5~Myr the best-fitting time-scale.
Dynamically equivalent models evolved for $\approx$ 3~Myr are outside
the error band (grey area), which indicates that
the dynamical processing of the binaries takes place during the first
Myr of cluster evolution.
This is because the ionization of soft
binaries takes a few initial crossing times, the cluster expands on this time-scale
because of the heating from the hard binaries mainly in 4-body interactions and
the mass loss due to relaxation, so that effectively the
binary population nearly freezes after a few initial crossing times.
Expansion from the compact radii to the equilibrium radii of open star clusters on the observed time-scale of about a dozen~Myr
is a result of the combination of the expulsion of residual gas given the small star-formation efficiencies and two-body relaxation 
and binary-star heating \citep{Gieles_2016,Megeath_2016,Banerjee_2017}.

We finish this section by discussing some caveats about the main assumptions in this work.
The initial cluster models are assumed here to be spherically symmetric and virialized,
which is a reasonable first-order approximation for embedded clusters.
Indeed, the time-scale for the formation of individual stars is $10^5\,$yr 
\citep[e.g.][]{Wuchterl_2003,Duarte_2013} while a cluster takes about 1~Myr to form most of its stars.
Once a star has largely accumulated its mass within 
about $10^5\,$yr it decouples from the hydrodynamical flows and becomes a ballistic 
particle in the cluster, such that at the onset of gas expulsion, after about 0.5--1~Myr,
the system is not far from dynamical equilibrium.
It tends to be also well mixed because the 
crossing times are typically $\lesssim$ 0.1~Myr, since embedded clusters are expected to
be born as compact structures \citep{Testi_1999,Marks_2012}.
Thus, although the proto-stars are in thin filamentary structures at birth 
\citep[e.g.][]{Konyves_2015,Hacar_2017}, these break up on a crossing time-scale 
and the stars virialise in the potential of the forming mass-segregated cluster \citep[e.g.][]{Bontemps_2010,Kirk_2011}.
%
%

However, asymmetries, substructures, gas clumps and deviations from virial equilibrium are found in most observed young star clusters 
in the nearby Universe. 
The dynamical processing with sub-structured initial conditions needs to be further studied, but observed very young clusters are too compact and too young to have been
formed from significant merging of many initially independent sub-clusters \citep{Banerjee_2015b,Banerjee_2015}. 


Even though the Monte Carlo code applied in this work is not designed for investigations of embedded clusters, it is still
a good computing machinery, since it is a reasonable first-order approximation. We note that involving more realistic dynamical simulations 
will be useful to further test the results achieved in this work.
In addition, solutions found under the present assumption of spherical symmetry should be 
dynamically equivalent to non-spherically symmetric initial conditions, and this is also
an important confirmation to be reached.

\section{CONCLUSIONS}
\label{conclusions}

We explored the concept of dynamical equivalence using Monte Carlo 
numerical simulations and compared them with observational properties
of Galactic binaries and embedded clusters. We show that this a
useful concept in explaining how the Galactic field is populated
via the dissolution of embedded clusters. We show that, for the
time-scale ($\lesssim$ 1~Myr) during which we expect dynamical
evolution to play a significant role in shaping the binary population,
there is a clear correlation between initial cluster mass and density/radius
(even though that associated with the radius is weaker),
which is consistent with that derived from observations, i.e. 
$\left( r_h/{\rm pc} \right)\,=\,0.1^{+0.07}_{-0.04} \; \left( M_{\rm ecl}/{\rm M_\odot}\right)^{0.13 \pm 0.04}$ \citep{Marks_2012}.
Our results indicate that, in fact, dynamical processing of the
binary population takes place for a time $\lesssim$ 1~Myr, which is consistent
with the time-scale associated with gas embedded star clusters \citep[e.g.][]{Wood_1989,Churchwell_2002,Churchwell_2010,Banerjee_2015b,Banerjee_2015,Banerjee_2017,Brinkmann_2017}.
Finally, our set of dynamically
equivalent models indicates that embedded clusters with different properties
contribute in populating the Galactic field and, in turn, the Galactic field 
population comes from different embedded clusters. This verifies the analytical
results arrived at by \citet{Marks_2011b}.

\section*{Acknowledgements}

We would like to thank an anonymous referee for the comments and suggestions that 
helped to improve this manuscript.
DB was supported by the CAPES Foundation, Brazilian Ministry of Education through the grant BEX 13514/13-0 and by the National Science Centre, Poland, through the grant UMO-2016/21/N/ST9/02938. 
PK thanks ESO, Garching, for support through a science visitor position. 
MG was partially supported by the National Science Centre, Poland, through the grant UMO-2016/23/B/ST9/02732.

\bibliographystyle{mnras}
\bibliography{references}

\bsp

\label{lastpage}

\end{document}